\documentclass[twocolumn,showpacs,aps]{revtex4}
\usepackage{epsf}
\def\R{{\rm I\hspace{-0.4ex}R}}

\def\curl{{\hbox{curl }}}
\def\div{{\hbox{div }  }}
\def\hb{\hbar}

\def\bO{{\bf \Omega}}
\def\tO{{\bf {\tilde \Omega}}}
\def\ep{\varepsilon}
\def\lep{{|\mathrm{log }\ \ep|}}
\def\ln{{\mathrm{log }}}

\def\dist{{\rm dist}}

\def\nab{\nabla}

\def\O{\Omega}

\def\a0{\alpha_0}
\def\a{\alpha}
\def\b{\beta}
\def\g{\gamma}
\def\dg{{\bf \delta}_\g}
\def\vdg{{\vec{\bf \delta}}_\g}
\def\rtf{\rho_{\hbox{\tiny TF}}}
\def\ortf{{\overline {\rho}}_{\hbox{\tiny TF}}}
\def\r0{\rho_{0}}
\def\too{\tilde\Omega}
\def\la{\lambda}
\def\be{\begin{equation}}
\def\ee{\end{equation}}
\def\beq{\begin{equation}}
\def\eeq{\end{equation}}

\def\cd{{\cal D}}

\begin{document}

\title{Vortex energy and vortex bending for a rotating Bose-Einstein condensate}

\author{Amandine Aftalion}
\email{aftalion@ann.jussieu.fr}
\affiliation{Laboratoire d'analyse num\'erique,
B.C.187,  Universit\'e Paris 6, 175 rue du Chevaleret, 75013 Paris,
France.}
\author{Tristan Riviere}
\email{riviere@math.polytechnique.fr}
\affiliation{Centre de Math\'ematiques, Ecole polytechnique,
 91128 Palaiseau Cedex, France.}
\date{\today}


\begin{abstract}
For a Bose-Einstein condensate placed in a rotating trap, we give a simplified expression of the Gross-Pitaevskii
energy in the Thomas Fermi regime, which only depends on the number and shape of the vortex lines. 
 Then we check numerically that when there is one vortex line, our simplified expression leads to solutions with a bent vortex  for a range of rotationnal velocities and trap parameters which are consistent with the experiments. 
\end{abstract}

\maketitle

\section{Introduction}

Since the experimental achievement of Bose-Einstein condensates in
confined alkali-metal gases in 1995, there has been a huge 
experimental and theoretical interest in these systems \cite{BuR,CD,DGPS,mal,FCS,FCS2,MCWD2,FS,MCWD,SF}. The study of vortices is one of the key issues. Two different groups have obtained vortices experimentally, the JILA group \cite{mal} and the ENS group \cite{MCWD2,MCWD}. In the ENS experiment, a laser beam is imposed on the magnetic trap holding the atoms to create a harmonic anisotropic rotating potential. For sufficiently large angular velocities,  vortices are detected in the system. Experimentally,
 the ENS group \cite{MCWD} has observed that when the vortex is nucleated,
 the contrast is not $100\%$ 
 which means that the vortex line is not straight but bending. Numerical computations solving the Gross-Pitaevskii equation \cite{C,GP1} have shown that there is a range of velocities for which the vortex line is indeed bending. The aim of this paper is to justify these observations theoretically in the Thomas Fermi regime. 
 We define an asymptotic parameter which is small in the Thomas Fermi regime and approximate the Gross-Pitaevskii energy  to obtain a simpler form of the energy which only depends on the shape of the vortex lines. Then we check numerically that our characterization leads to solution with a bent vortex for a range of  values of the rotationnal velocity which are consistent with the ones obtained numerically \cite{C}. Let us point out that Svidzinsky and Fetter \cite{SF2} have studied the dynamics of a vortex line depending on its curvature. For a vortex velocity equal to 0, the equation obtained in \cite{SF2} is the same as the equation corresponding to the minimum of our approximate energy, though the  formulation in \cite{SF2} was not derived from energy considerations. Moreover, their analysis is only valid for a single vortex line.

The Gross-Pitaevskii energy
 provides a very good description
of Bose-Einstein condensates: it is assumed that the $N$ particles of
the gas are condensed in the same state for which the wave function
$\phi$ minimizes the Gross-Pitaevskii energy. In the ENS experiment, a laser is applied to the trap which makes it rotate.
By introducing a rotating frame at the angular
velocity $\tO = \too{\bf e}_z$,  the trapping potential becomes time
independent, and the wave function $\phi$ minimizes the energy
\begin{eqnarray}
\label{BE}
{\cal E}_{3D}(\phi) &=& \int {\hb^2 \over{2m}} |\nabla \phi|^2+{{\hb {\tO}}}\cdot
(i\phi, \nabla \phi\times {\bf x}) \nonumber\\
&&+{m\over 2}
 \sum_{\alpha} \omega^2_\alpha  r^2_\alpha |\phi|^2 +{N\over 2} g_{3D}|\phi|^4 , 
\end{eqnarray}
under the constraint $\int |\phi|^2 =1$. Here,
for any complex quantities $u$ and $v$ and their
complex conjugates $\bar{u}$ and $\bar{v}$, 
 $(u,v)=(u\bar{v}+\bar{u}v)/2$.
  
We want to nondimensionnalize the energy in order to get a  parameter which is small in the Thomas-Fermi regime. This framework of study has been developed by one of the authors in \cite{AD}, except that \cite{AD} was a two dimensionnal  study for a condensate confined in the $z$ axis.
 We define the characteristic length  $d=(\hb / m\omega_x)^{1/2}$ and assume
 $\omega_y=\a\omega_x$, $\omega_z=\b\omega_x$.
  We set
$$\ep^2\sqrt{\ep}= {{\hb^2d}\over {2Ngm}}={{d}\over {4\pi Na}},$$ 
where $g_{3D}=4\pi\hb ^2a/m$.  For numerical applications, we are going to use the experimental values of the ENS group \cite{C,MCWD},   $m=1.445.10^{-26}kg$, $a=5.8.10^{-11}m$, $N=1.4.10^{5}$ and $\omega_x=1094s^{-1}$ with $\a=1.06$, $\b=0.067$. We find that $\ep=0.0174$,
 thus, $\ep$ is
small, which will be our asymptotic regime. 
We re-scale the distance by $R=d/\sqrt \ep$ and
define $u({\bf r})=R^{3/2} \phi({\bf x})$ where ${\bf x}=R{\bf r}$ and we set $\O =\too /\ep\omega_x$.
The velocity $\O$ is chosen such that $\O <1 / \ep$, that is the trapping potential is stronger than the inertial potential. The energy can be rewritten as:
\begin{eqnarray}
\label{be3d}
{E}_{3D}(u) &=&\int {1\over 2} |\nabla u|^2+ \bO \cdot (iu,\nabla
u\times {\bf r})\\&&+{1\over
2\ep^2}(x^2+\a^2y^2+\b^2z^2)|u|^2+{1\over 4\ep^2}|u|^4\nonumber
 \; .
\end{eqnarray}
Due to the constraint $\int |u|^2 =1$,  we can add to ${ E}_{3D}$
any multiple of 
$\int
|u|^2$ so that it is equivalent to minimize 
$${1\over 2}\int |\nabla u |^2+ \bO \cdot (iu,\nabla u\times {\bf r})+{1\over
4\ep^2}|u|^4-{1\over {2\ep^2}}\rtf({\bf r})|u|^2$$ 
where $\rtf({\bf r})=\r0 -(x^2+\a^2y^2+\b^2z^2)$ for some constant $\r0$ 
to be determined.  Let $\cd$ be the ellipse
$\{\rtf >0\}=\{x^2+\a^2 y^2+\b^2z^2<\r0\}$. We impose the following constraint
on  $\rtf$:
\beq\label{cons}\int_{\cd}\rtf({\bf r})=1.\eeq
Indeed, as $\ep$ tends to 0, the minimizer will satisfy that $|u|^2$ will
be close to $\rtf$ so that the constraint will be satisfied automatically by $u$ if we impose (\ref{cons}). In other words, $\rtf$ is the Thomas Fermi approximation of $u$. Equation
 (\ref{cons})
leads to 
\beq\label{rho0}
\r0^{5/2}=15\a\b /8 \pi.\eeq 
To study the problem analytically, it is reasonable to minimize
the energy over the domain $\cd$ with zero boundary data for
$u$. Indeed, when $\rtf\leq 0$,  the energy is convex so that the
minimizer $u$ goes to zero exponentially at infinity (see  the numerical
observation in \cite{CD} and the analysis on the behavior near the
boundary of $\cd$  as well as the decay at infinity of the order
parameter in \cite{DPS,FF}). 
 We  consider  the problem
$$\min E_\ep(u)\; \;\hbox{subject to }\;
u\in H_0^1(\cd),\;  \int_{\cd} |u|^2=1\; 
\eqno{(P)}$$
where
\begin{eqnarray}
\label{be}
  E_\ep (u)&=&\int_\cd {1\over 2}|\nabla u |^2+  \bO \cdot (iu,\nabla u\times {\bf
r})\nonumber\\
&& +{1\over 4\ep^2}(\rtf({\bf r})-|u|^2)^2 .
\end{eqnarray} 
Note that a critical point $u$ of $E_\ep$ is a solution of
\beq\label{eqbe}
-\Delta u+2i(\bO\times {\bf r}).\nabla u={1\over \ep^2}u(\rtf-|u|^2)+\mu_\ep u 
 \hbox{ in }  \cd,
\eeq
with $u=0$ on $\partial \cd$ and $\mu_\ep$ is the Lagrange multiplier. The specific choice of $\r0$ in (\ref{rho0}) will imply that the term $\mu_\ep u$ is negligible in front of $\rtf u/\ep^2$.

We have set the framework of study of our energy. In Section 2, we will make an asymptotic development of the energy taking into account that $\ep$ is small (but $\lep$ is not big). Then in Section 3, we will check that our approximate energy yields a solution which is consistent with the numerical and experimental observations.

\section{Asymptotic development of the energy}
Our aim is to decouple the energy into 3 terms: a part coming from the solution without vortices, a vortex
contribution and a term due to rotation. 

\subsection{The solution without vortices}
Firstly, we are interested in solutions without vortices, that is $u$
has no zero in the interior of $\cd$. 
Thus we consider functions of the form $\eta=fe^{iS}$,
where $\eta$ is in $H_0^1(\cd)$ and $f$ is real and has no zero in
the interior of $\cd$. We consider first minimizing $E_\ep$ over
such functions without imposing the
constraint that the $L^2$ norm is 1,
that is, $f$ and $S$ minimize
\begin{eqnarray}\label{fSener}
{\cal E}_\ep(f,S)&=&\int_{\cd} {1\over 2}|\nabla f |^2+{1\over
4\ep^2}(\rtf-f^2)^2\nonumber\\
&&+{1\over 2}\int f^2|\nabla S-\bO\times {\bf r}|^2-f^2\O^2r^2, 
\end{eqnarray}
where ${\bf r}=x{\bf e}_{x}+y{\bf e}_y$.
We have $f_\ep=0$ on $\partial \cd$ and 
\beq\label{f}
-\Delta f_\ep+f\nabla S_\ep(\nabla S_\ep-2\bO\times {\bf r})={1\over \ep^2}f_\ep(\rtf-f_\ep^2)
\hbox{ in } \cd , 
\eeq
\beq\label{S}
\div (f_\ep^2(\nabla S_\ep-\bO\times {\bf r}))=0.\eeq
Equation (\ref{S}) implies that there exists $\xi_\ep$ in $H^2(\cd)\cap
H^1_0(\cd)$ such that 
\beq\label{xinp}
f_\ep^2(\nabla S_\ep-\bO\times {\bf r})=\O\ \curl\ \xi_\ep. \eeq
 So $\xi_\ep$ is the
unique solution of 
\beq\label{xieq}
\curl ( {1\over f_\ep^2} \curl\xi_\ep)=-2 \hbox{ in } \cd, \quad \xi_\ep=0 \hbox{ on
} \partial \cd.\eeq 
In the special case where the cross section of $\cd$ is a disc, the minimum of
(\ref{fSener}) is reached for $\nabla S=0$ but this is not the case if
 the cross section is an ellipse and there is a non trivial solution of (\ref{S}). 
 As $\ep$ tends to 0, since the ellipticity of the cross-section is small, $f_\ep^2$ tends to $\rtf$ in every compact subset of $\cd$ and
the function $\xi_\ep$ given by (\ref{xinp}) or (\ref{xieq})  tends to
the unique solution $\xi$ of  
\beq\label{xilim}
\curl({1\over \rtf}\curl \xi)=-2 \hbox{ in } \cd, \quad \xi=0 \hbox{ on }
\partial \cd.\eeq 
One can easily get that
\beq\label{valxi} 
\xi(x,y)={-\rtf^2(x,y)/(2+2\a^2)}{\bf e}_z.\eeq
Using (\ref{xinp}), we can define
 $S_0$, the limit of $S_\ep$,
 to be the solution of  $\rtf(\nabla S_0-\bO\times{\bf r})=\O\ \curl \xi$
with zero value at the origin. We have $S_0=C\O xy$ with
$C=(\a^2-1)/(\a^2+1)$. We see that $S_0$ vanishes when $\a=1$ that is when the cross-section is a  disc.   This computation is consistent with the one in
\cite{SF}, though it is derived in a different way.
\subsection{Decoupling the energy}
Let $\eta_\ep=f_\ep e^{iS_\ep}$ be the vortex free 
minimizer of $E_\ep$ discussed previously without imposing
the constraint on the norm of $u$. Let
$u_\ep$ be a configuration that will minimize $E_\ep$
and  let $v_\ep=u_\ep/\eta_\ep$. Since $\eta_\ep$  satisfies the Gross
Pitaevskii equation (\ref{f})-(\ref{S}), we have 
\begin{eqnarray}
&&\int_{\cd} (|v_\ep|^2-1) (-{1\over 2}\Delta f_\ep^2-{1\over \ep^2}f_\ep^2
(\rtf - f_\ep^2)+|\nabla f_\ep e^{iS_\ep}  |^2\nonumber\\
&&\quad\quad-2f_\ep ^2(\nab S_\ep\cdot
\bO\times {\bf r}))=0.\nonumber
\end{eqnarray}
Using this identity, 
one can get that the energy $E_\ep(u_\ep)$ decouples as follows
\beq
\label{splitener}
E_\ep(u_\ep)=E_\ep(\eta_\ep)+G_{\eta_\ep}(v_\ep)+I_{\eta_\ep}(v_\ep)
\eeq
where
$$G_{\eta_\ep}(v_\ep)=\int_{\cd} {1\over 2}|\eta_\ep|^2|\nabla
v_\ep|^2+{{|\eta_\ep|^4}\over{4\ep^2}}(1-|v_\ep|^2)^2,$$ 
and
$$I_{\eta_\ep}(v_\ep)=\int_{\cd} |\eta_\ep|^2 (\nab S_\ep - \bO\times {\bf r}) \cdot
(iv_\ep,\nabla v_\ep).$$
The first term in the energy is independent of the solution $u_\ep$, so we have to compute the next two and find for which configuration $u_\ep$ the minimum is achieved. We assume that the solution $u_\ep$ has a vortex line along $\g$ and we call $\dg$ the dirac measure along the line. Our aim is to estimate the energy of $u_\ep$ depending on $\g$.
 A first approximation consists in writing that $|\eta_\ep|^2$ tends to $\rtf$ when $\ep$ is small so that we can approximate $G_{\eta_\ep}$ by $G_{\sqrt\rtf}=G_\ep$ and $I_{\eta_\ep}$ by $I_{\sqrt\rtf}=I_\ep$.

\subsection{Estimate of $G_\ep(v_\ep)$}
We want to estimate
$$G_{\ep}(v_\ep)=\int_{\cd} {1\over 2}|\rtf|^2|\nabla
v_\ep|^2+{{|\rtf|^4}\over{4\ep^2}}(1-|v_\ep|^2)^2.$$ 
The mathematical techniques to approximate $G_\ep$ have been introduced by \cite{BBH} in the 2 dimensionnal case and by \cite{R} in dimension 3, when $\ep$ is very small. The problem here is that $\ep=0.0174$ so that $\lep$ is not large and there will be additional terms in the asymptotic expansion.
 For a minimizing configuration, one can prove that $v_\ep$ is tending to 1 everywhere except close to the vortex line $\g$.
We define
\beq\label{T}
T_{\la\ep}=\{ x\in \cd \hbox{ s.t. } \dist(x,\g)\leq\la\ep\},\eeq
and assume that $\la\ep$ is small, $\la$ being a nondimensionnal parameter to be fixed later on.
Then we split $G_{\ep}$ into two integrals: one in $T_{\la\ep}$ and the other in $\cd\setminus T_{\la\ep}$.
\subsubsection{Estimate near the vortex core}
We are going to estimate $G_\ep$ in $T_{\la\ep}$.
At each point $\g(t)$ of $\g$, we define $\Pi^{-1}(\g(t))$ to be the plane orthogonal to $\g$ at $\g(t)$. Since $\la\ep$ is small, we assume that $\rtf$ is constant in $\Pi^{-1}(\g(t))\cap T_{\la\ep}$ and we call the value $\rho_t=\rtf(\g(t))$.
We want to compute
\begin{eqnarray*}
\int_{T_{\la\ep}} {1\over 2}\rtf |\nabla v_\ep|^2+{{\rtf^2}\over {4\ep^2}}(1-|v_\ep|^2)^2\\\simeq \int_\g {{\rho_t}\over 2} \int_{\Pi^{-1}(\g(t))\cap T_{\la\ep}}|\nabla v_\ep|^2+{{\rho_t}\over {2\ep^2}}(1-|v_\ep|^2)^2.
\end{eqnarray*}
Since $u_\ep$ is a minimizing configuration of $E_\ep$, after scaling by $r\sqrt{\rho_t}/\ep$, we find that 
 $v_\ep$ is very close to $u_1(r\sqrt{\rho_t}/\ep)$, where
 $u_1(r,\theta)=f_1(r)e^{i\theta}$ is the solution with a single zero at the origin of
$$\Delta u+u(1-|u|^2)=0\ \hbox{in } \R^2.$$
Thus,
\begin{eqnarray*}
&&\int_{\Pi^{-1}(\g(t))\cap T_{\la\ep}}|\nabla v_\ep|^2+{{\rho_t}\over {2\ep^2}}(1-|v_\ep|^2)^2\\
\simeq && \int_{B_{\la\ep}} \Bigl | \nabla \Bigl (f_1(r\sqrt{{\rtf}\over \ep^2}\Bigr ) e^{i\theta}\Bigr ) \Bigr | ^2+{{\rho_t}\over {2\ep^2}}\Bigl ( 1-f_1^2\Bigl ( r\sqrt{{\rtf}\over \ep^2}\Bigr ) \Bigr )^2
\\ =&&\int_{B_{\la\sqrt{\rho_t}}} |\nabla u_1|^2+{1\over 2}(1-|u_1|^2)^2\\
\simeq &&c_*+2\pi\ln({\la\sqrt{\rho_t}}),
\end{eqnarray*}
where 
$$c_*=\int_{\R^2}{f_1'}^2+{1\over 2}(1-f_1^2)^2+\int_{\R^2\setminus{B_1}}{{f_1^2-1}\over r^2} +\int_{B_1}{{f_1^2}\over r^2}.$$
The last approximation is good if $\la\sqrt{\rho_t}$ is large (in fact bigger than 3 is enough). The existence of $\la$ is justified by the fact that $\sqrt{\rtf}/\ep$ is much bigger than 1, except very close to the boundary.

The final estimate of this section is
\beq \label{Gint}
G_\ep(v_\ep)_{|{T_{\la\ep}}}\simeq\int_\g \rtf({{c_*}\over 2}+\pi\ln({\la\sqrt{\rtf}})) dl
\eeq
\subsubsection{Estimate away from the vortex core}
We are going to estimate $G_\ep$ in $\cd\setminus T_{\la\ep}$. In this region $v_\ep \simeq 1$, so that only the kinetic energy of the phase has a contribution.
\begin{eqnarray*}
&&\int_{\cd\setminus T_{\la\ep}} {1\over 2}\rtf |\nabla v_\ep|^2+{{\rtf^2}\over {4\ep^2}}(1-|v_\ep|^2)^2\\
&&\simeq \int_{\cd\setminus T_{\la\ep}}{1\over 2}\rtf |\nabla \phi_\ep|^2,
\end{eqnarray*}
where $\phi_\ep$ is the phase of $v_\ep$. Of course, $\phi_\ep$ is not defined everywhere. But let $\psi$ be such that $\div \psi=0$ and
$$\curl \psi=\rtf\nabla\phi.$$
Then $\psi$ is the unique solution of
\beq\label{psi}
\curl \Bigl ( {1\over {\rtf}}\curl \psi\Bigr )=2\pi\vdg ,\ \psi=0\hbox{ on }\partial \cd,
\eeq
where $\vdg$ is the vectorial dirac measure along $\g$
and
\begin{eqnarray*}
\int_{\cd\setminus T_{\la\ep}}{1\over 2}\rtf |\nabla \phi_\ep|^2
=\int_{\cd\setminus T_{\la\ep}}{1\over {2\rtf}} |\curl \psi|^2\\
=-{1\over 2}\int_{\partial T_{\la\ep}}\psi\cdot \nabla \phi_\ep\times \nu
\end{eqnarray*}
where $\nu$ is the outward unit normal. We will see that $\psi$ is almost constant at a distance $\la\ep$ from $\g$ and we call this value $\psi_{\la\ep}(\g)$.
Since the vortex line has a winding number $2\pi$, 
$$\int_{\cd\setminus T_{\la\ep}}{1\over 2}\rtf |\nabla \phi_\ep|^2\simeq \pi \int_\g \psi_{\la\ep}(\g)\cdot dl.$$
We have to compute $\psi$ on $\partial T_{\la\ep}$. The computation is inspired by the paper of Svidzinsky and Fetter \cite{SF2}.
 It follows from (\ref{psi}) that $\psi$ satisfies
$$-\Delta \psi-{{\nabla \rtf}\over \rtf}\times \curl \psi=2\pi\rtf\vdg .$$
Let $x_0\in\g$. We denote by ${\bf e}_3=\dot{\g}(x_0)$ and $({\bf e}_1,{\bf e}_2,{\bf e}_3)$ an orthogonal base in local coordinates. Then $\psi$ has coordinates $\psi_i$ in ${\bf e}_i$ and the variations of $\psi_3$ are the only ones of influence in the equation for $\psi$, since we want to compute $\psi\cdot dl$. We approximate the equation for $\psi$ by
\beq\label{psi3}
-\Delta \psi_3+{{\nabla \ortf}\over \ortf}\cdot \nabla \psi_3=2\pi\rtf\dg ,
\eeq
where $\ortf(x^1,x^2)=\rtf(x^1,x^2,x^3_0)$.
 Let $\Xi=\psi_3/\sqrt{\ortf}$. Then it follows from (\ref{psi3}) that $\Xi$ satisfies
\beq\label{Xi}
-\Delta \Xi+\mu \Xi=2\pi\sqrt{\rtf}\dg
\eeq
where
\beq\label{mu}
\mu=\sqrt{\ortf}\Delta {1\over{\sqrt{\ortf}}}=\sqrt{\rtf}\Delta _\perp{1\over{\sqrt{\rtf}}}.
\eeq
Here $\Delta_\perp$ is the laplacian in the plane perpendicular to ${\bf e}_3=\dot{\g}(x_0)$. If the cross-section is a disc one can compute $\mu$. We denote by $\theta$ the angle of ${\bf e}_3$ that is ${\bf e}_3=\cos\theta {\bf e}_r+\sin\theta{\bf e}_z$and $(r,z)$ are the coordinates of $x_0$ in the original frame. Then
\beq\label{mudisc}
\mu={{(1+\sin^2\theta)+\b^2\cos^2\theta}\over \rtf}+{{3(r\sin\theta-\b^2z\cos\theta)^2}\over \rtf^2}.
\eeq
Note that $\mu>0$. In fact our numerical computations even yield $\mu>7$. Our aim is now to give an approximate expression for $\Xi$.  We locally approximate the curve $\g$ near the point $x_0$ by the parabola $x=kz^2/2$, where $k$ is the curvature of $\g$ at $x_0$. This is where we use the same ideas as in \cite{SF2}. Note that in our approximations, we are only taking into account the shape of $\g$ close to $x_0$. The justification for this relies on the fact that $\mu>7$ as our numerics show. Indeed if we solve
$$-\Delta X+\mu X=f$$
where $f$ is supported at a distance $d$ of $x_0$. Then using the Green function, we find that 
$$|X|\leq{{e^{-\sqrt \mu d}}\over{4\pi \mu^3d}}.$$
In particular, for $d=0.1$, this gives an error less than $10^{-3}$. This is to be compared to the Euler constant and our approximation is reasonnable. We rewrite (\ref{Xi}) in local coordinates to get
$$-\Delta\Xi+k\partial_{x_1}\Xi+\mu\Xi=2\pi\sqrt{\rtf(x_0)}\delta_{{\bf e}_3},$$
where $\delta_{{\bf e}_3}$ is the dirac mass supported along the line ${\bf e}_3$. Thus
\beq\label{Xiexp}
-\Delta(e^{{-kx_1}\over 2}\Xi)+\Bigl (\Bigl ( {k \over 2}\Bigr )^2+\mu\Bigr )
\Bigl (e^{{-kx_1}\over 2}\Xi  \Bigr )=2\pi\sqrt{\rtf(x_0)}\delta_{{\bf e}_3}.
\eeq
The solution of this equation is
$$\sqrt{\rtf(x_0)}K_0\Bigl ( \sqrt{\mu+{k^2\over 4}}\dist (x,\g) \Bigr ), $$
where $K_0$ is a modified Bessel function. In particular, $K_0(x)\simeq -\ln(e^{C_0}x/2)$ for small $x$ where $C_0\simeq 0.577$ is the Euler constant. Hence, we deduce
\beq\label{estpsi}
\psi(x)\simeq -\rtf \ln\Bigl ( {{e^{C_0}}\over 2}\sqrt{\mu+{k^2\over 4}}\dist (x,\g)\Bigr )\dot{\g}.
\eeq

Thus we conclude by the estimate for $G_\ep(v_\ep)$ in $\cd\setminus T_{\la\ep}$
\beq\label{Gout}
G_\ep(v_\ep)_{ | \cd\setminus T_{\la\ep}}\simeq -\pi\int_\g \rtf\ln\Bigl ( {{e^{C_0}}\over 2}\sqrt{\mu+{k^2\over 4}}\la\ep\Bigr )\ dl.
\eeq

\subsection{Estimate of $I_\ep(v_\ep)$}
We want to estimate 
\beq\label{I}
I_{\ep}(v_\ep)=\int_{\cd} \rtf (\nab S_\ep - \bO\times {\bf r}) \cdot
(iv_\ep,\nabla v_\ep).\eeq
Recall that the unique solution of (\ref{xieq}) satisfies $\rtf(\nabla S_\ep-\bO\times{\bf r})=\O\ \curl \xi_\ep$. Hence we integrate by part in (\ref{I}) to get
$$I_{\ep}(v_\ep)=\O\int_{\cd}\xi_\ep\cdot \curl(iv_\ep,\nabla v_\ep).$$
Let $\phi_\ep$ be the phase of $v_\ep$.
 Since $v_\ep$ is tending to one everywhere except on the vortex line, 
 then  $(iv_\ep,\nabla v_\ep)\sim \nabla\phi_\ep$, hence 
we can approximate $\curl(iv_\ep,\nabla v_\ep)$ by $2\pi\vdg$. We use the value of $\xi$ given by (\ref{valxi})  and the fact that $\dot\g(t)\cdot {\bf e}_z=dz$ to get
\beq\label{Iapp}
I_{\ep}(v_\ep)\simeq -{{\O\pi}\over {(1+\a^2)}}\int_\g \rtf^2\ dz.
\eeq

\subsection{Final estimate for the energy}
We use (\ref{splitener})-(\ref{Gint})-(\ref{Gout})-(\ref{Iapp}) to derive the energy of a solution with a vortex line. Indeed  the energy of any solution minus the energy of a solution without vortex is roughly the vortex contribution in the sense:
\beq
E_\ep(u_\ep)-E_\ep(\eta_\ep)
\simeq {\cal E}_\g.
\eeq
We find that the vortex contribution ${\cal E}_\g$ is
\begin{eqnarray}\label{enerapp}
{\cal E}_\g=&&\int_\g \rtf \Bigl ({c_*\over 2}+\pi \ln\Bigl ( {2\over{\ep e^{C_0}}}\sqrt{{\rtf }\over {\mu+{k^2\over 4}}}\Bigr )\Bigr ) \ dl\nonumber\\
&&-{{\O\pi}\over {(1+\a^2)}}\int_\g \rtf^2 \ dz.
\end{eqnarray}
 Hence if the right-hand-side of (\ref{enerapp}) is negative, it means that it is energetically favorable to have vortices. Note that in the first integral of ${\cal E}_\g$, we have $dl=|\dot{\g}(z)|dz$ whereas in the second one, we have $dz$.

If the vortex line is straight, our computation yields
\begin{eqnarray}\label{vdroit}
{{\r0^{3/2}}\over\b}&&\Bigl ( {2\over 3}\Bigl ({c_*\over 2}+\pi \ln\Bigl ( {{\sqrt 2}\over{\ep e^{C_0}}}\Bigr )\Bigr )+{{2\pi}\over 3}\ln\r0\nonumber \\
&&+\pi\Bigl ( {{-10}\over 9}+{4\over 3}\ln 2\Bigr )
-\O{{8\pi\r0}\over{15(1+\a^2)}}\Bigr )
.
\end{eqnarray}
This gives a critical angular velocity $\O_1$ for which a straight vortex has a lower energy than a vortex free solution. With our experimental data, it yields $\O_1\sim 22.45$, that is $\tilde{\O}_1/\omega_x\sim 0.39$. We are going to see in the numerical section that for $\O<\O_1$, a bent vortex has a negative energy.

\subsection{Case of several vortices}
Let us assume that the solution $u_\ep$ has $n$ vortices along the lines $\g_i$, $1\leq i\leq n$. We want to estimate the energy in this case. For each $\g_i$, we define $T_{i,\la\ep}$ as in (\ref{T}).

One can check that the estimates (\ref{Iapp}) and (\ref{Gint}), respectively for $I_\ep(v_\ep)$ and for $G_\ep(v_\ep)$ close to each vortex core, are unchanged if the integral along $\g$ is replaced by the sum of the integrals along $\g_i$. The only difference is for the estimate away from the vortex cores where we have to take into account the interaction between the vortex lines. Let us denote $\cd_n=\cd\setminus \cup_j T_{j,\la\ep}$. We still have
\beq \label{Gm}
G_\ep(v_\ep)_{|{\cd_n}}\simeq
\int_{\cd_n}{1\over {2\rtf}} |\curl \psi|^2,
\eeq
where $\psi=\sum_i \psi_i$ and $\psi_i$ solves (\ref{psi}) with $\g_i$ instead of $\g$. Thus, we need to estimate
\beq\label{enmul}
\sum_i \int_{\cd_n}{1\over {2\rtf}} |\curl \psi_i|^2+\sum_{i\neq k} \int_{\cd_n}{1\over {2\rtf}} |\curl \psi_k|\cdot |\curl \psi_i| .
\eeq
The first integral is estimated as in section C.2 by
\beq\label{sum1}
\sum_i  -\pi\int_{\g_i} \rtf\ln\Bigl ( {{e^{C_0}}\over 2}\sqrt{\mu+{k^2\over 4}}\la\ep\Bigr )\ dl.
\eeq
As for the second integral in (\ref{enmul}), we integrate it by part to get
\beq\label{sum2}
\pi\sum_{i\neq k}\int_{\g_i}\psi_{k}\cdot dl.
\eeq
The computation of $\psi_k(x)$ from section C.2 is still valid and we have
 $\psi_k(x) \simeq -\rtf K_0(\sqrt{\mu+{k^2\over 4}}dist(x,\g_k))$.
 This yields the contribution of $n$ vortex lines (to be compared with (\ref{enerapp}) for 1 vortex)
\begin{eqnarray}\label{enerappn}
{\cal E}_n=&&\sum_i\int_{\g_i} \rtf \Bigl ({c_*\over 2}+\pi \ln\Bigl ( {2\over{\ep e^{C_0}}}\sqrt{{\rtf }\over {\mu+{k^2\over 4}}}\Bigr )\Bigr ) \ dl\nonumber\\
&&-{{\O\pi}\over {(1+\a^2)}}\int_{\g_i} \rtf^2 \ dz\\
&&-\pi\sum_{i\neq k}\int_{\g_i}\rtf K_0(\sqrt{\mu+{k^2\over 4}}dist(x,\g_k)) dl,
\end{eqnarray}
where $K_0$ is a modified Bessel function. Note that
 the curves are going to interact only in the region where they are close to one another.

\section{Comparison with the numerics}
We are interested in the shape of the vortex line that minimizes (\ref{enerapp}) according to the value of $\O$. We write $(\g(t)=r(t),z(t))$ and we  assume that the vortex line is in the plane $(y,z)$. 
We will denote $\rtf(t)=\r0-\a^2r^2(t)-\b^2z^2(t)$ and we define 
$$C(t)={c_*\over 2}+\pi \ln\Bigl ( {2\over{\ep e^{C_0}}}\sqrt{{\r0-\a^2r^2(t)-\b^2z^2(t)}\over {\mu+{k^2\over 4}}}\Bigr ).$$
Since (\ref{enerapp}) does not depend on the parametrization $\g(t)$,
we choose a special parametrization on the curve such that
\beq\label{crho1}
C^2(t)\rtf^2(t)(\dot{r}^2(t)+\dot{z}^2(t))=1.\eeq
Then it is equivalent to minimize
\beq\label{e2}
\int_\g C^2(t)\rtf^2(t)(\dot{r}^2(t)+\dot{z}^2(t)) dt-{{\O\pi}\over {(1+\a^2)}}\int_\g \rtf^2 (t)\dot{z}(t) dt\eeq
under the constraint (\ref{crho1}). In our computations below, we will proceed to a minimization of (\ref{e2}) releasing the constraint (\ref{crho1}). Indeed, computations show that (\ref{crho1}) is true from $t=0$ to $t^*$ where the shape of the vortex is determined.
 Under the assumption that $\mu$ and the curvature do not vary too much along the curve, we derive an equation for the minimum $\g$:
\begin{eqnarray*}
{d\over{dt}}(C^2\rtf^2\dot{r})=-{{2\a^2r(t)}\over \rtf(t)}+{{2\a^2\O}\over {(1+\a^2)}}\rtf r(t)\dot{z}(t),\\
{d\over{dt}}(C^2\rtf^2\dot{z})=-{{2\b^2z(t)}\over \rtf(t)}-{{2\a^2\O}\over {(1+\a^2)}}\rtf r(t)\dot{r}(t).
\end{eqnarray*}
Thus, we solve this system
with initial conditions $r(0)=r_0$, $\dot{r}_0=0$, $z(0)=0$, $C(0)\rtf(0)\dot{z}(0)=1$.

\begin{figure}[htb]
\vspace{0.1in}
\centerline{\epsfxsize=2.8in\epsfbox{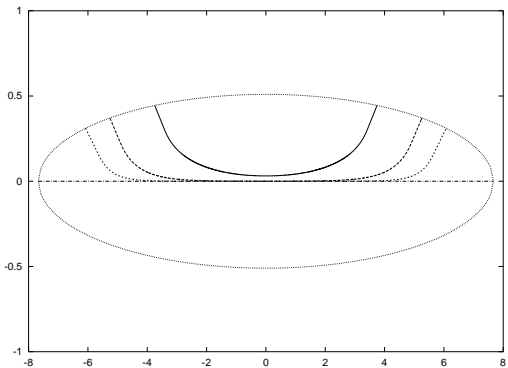}}
\caption{The vortex line for various values of $\Omega$ in the $z-y$ plane: $\O=21.8$ (straight line), $\O=25.8$ (dotted line), $\O=33.1$ (dashed line).}
\label{figvor}
\end{figure}

 We let $r_0$ vary in order to find the minimizing solution. We have   drawn the vortex line for the minimizing solution for some values of $\O$ in Figure \ref{figvor}. We find that indeed the vortex line is bending for a range of $\O$.
 The bent vortex  starts to exist near the boundary of the ellipse, that is $y=\sqrt\r0/\a$, $z=0$ for $\O_0=21.2$, that is $\tilde\O_0/\omega_x=0.368$. As $\O$ increases, the value of $r_0$ decreases: $r_0=0.03$ for $\O=21.8$, $r_0=2.9\ 10^{-4}$ for $\O=25.8$, $r_0=10^{-6}$ for $\O=33.1$. As $\O$ increases, $r_0$ becomes smaller, the bent vortex  gets very close to the straight vortex. The shape of the vortex lines are similar to those obtained in \cite{GP1} using the full Gross Pitaevskii energy.

\begin{figure}[htb]
\vspace{0.1in}
\centerline{\epsfxsize=2.8in\epsfbox{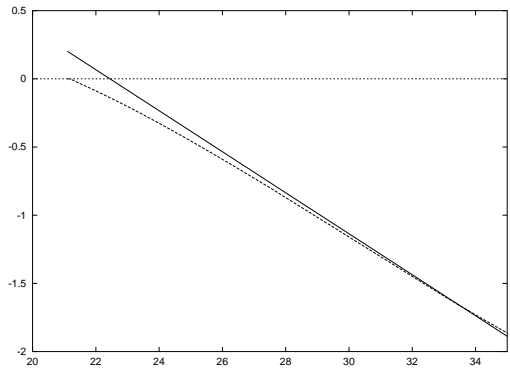}}
\caption{The energy vs. $\Omega$ curves for the solution with a straight vortex (solid line) and a bent vortex (dotted line).}
\label{figener}
\end{figure}

We plot the energy of the straight vortex line and the bent vortex  vs $\O$ in Figure \ref{figener}.  
 One can observe that for $\O_c=21.8$, that is $\tilde \O_c/\omega_x=0.38$ in the initial units, the energy of the bent vortex  starts to be negative (that is below the energy of a  solution without vortex), while the energy of a straight vortex line is positive. For $\O=33.1$, the energy of the bent vortex  and of a straight vortex line become equal. These results are consistent with the ones in \cite{C}. They obtain the same value of $\O_c$ for which the bent vortex has a negative energy.

Let us point out that the bent vortex is a minimizer even if the cross section is a disc. Nevertheless, when $\ep$ is fixed, if $\b$ gets too big, the straight vortex becomes the minimizer, which is the case for $\b=1$. Our analysis could give the critical value of $\beta$ above which the vortex line should be straight.

We believe that our analysis justifies why in the conditions of the ENS experiment, when a vortex is nucleated, the contrast is not 100$\%$ : indeed, a bent vortex has a lower energy than a straight vortex. Nevertheless, the velocity of nucleation in the experiment is higher than  our critical angular velocity $\O_c$: we compute the thermodynamical critical velocity whereas the velocity of nucleation is likely to be closer to the velocity where the vortex free solution loses local stability. Once the first vortex is obtained experimentally, if $\O$ is decreased, the bent vortex is likely to exist down to $\O_0$.

When there are several vortices, we were not able to find numerically the shape of the vortex lines which minimize (\ref{enerappn}) but we believe that our simplified energy is a good description of the experiments \cite{MCWD} and the numerics \cite{GP1} and we hope that it can be easier to handle than the full Gross Pitaevskii energy.

\section{Conclusion}
We have obtained a simplified expression (\ref{enerapp}) of the energy of a minimizing solution of the Gross Pitaevskii energy with a vortex line $\g$ and (\ref{enerappn}) for $n$ vortex lines $\g_i$. This expression depends on the shape of the vortex line. This has allowed us to draw the vortex line for the minimizing solution and compute its energy. We have seen that there is a range of rotationnal velocities for which a bent vortex line has a lower energy than a straight vortex and a vortex free solution. These computations on the simplified expression of the energy are in agreement with the computations on the full energy \cite{C,GP1}.

\begin{acknowledgments}
\vspace{-0.2in}
The authors would like to warmly thank Y.Castin for explaining the work of his team at the ENS and for very
interesting and encouraging discussions.
The authors are very indebted to him for his critics and remarks on the manuscript. The work has also largely benefited from discussions with E.Sandier and S.Serfaty. The authors wish  to thank them very much.
 
\end{acknowledgments}

\end{document}